\documentclass[reqno,11pt]{amsart}
\usepackage{multind}
\usepackage{hyperref}
\usepackage{graphicx}
\usepackage{amscd}
\usepackage{slashed}
\usepackage{amssymb}
\usepackage[mathscr]{eucal}
\numberwithin{equation}{section}
\allowdisplaybreaks[1]

\title[On Minimizers of Causal Variational Principles]{On the Structure of Minimizers of \\
Causal Variational Principles in the \\
Non-Compact and Equivariant Settings}

\author[Y.\ Bernard]{Yann Bernard}
\address{Mathematisches Institut \\ Universit\"at Freiburg \\ D-79104 Freiburg \\ Germany}
\email{yann.bernard@math.uni-freiburg.de}

\author[F.\ Finster]{Felix Finster \\ \\ May 2012}
\address{Fakult\"at f\"ur Mathematik \\ Universit\"at Regensburg \\ D-93040 Regensburg \\ Germany}
\email{finster@ur.de}
\thanks{Both authors are supported by the Deutsche Forschungsgemeinschaft (the first author by the DFG SFB 71 fund).}

\newtheorem{Def}{Definition}[section]
\newtheorem{Thm}[Def]{Theorem}
\newtheorem{Prp}[Def]{Proposition}
\newtheorem{Lemma}[Def]{Lemma}

\newcommand{\Thanks}{\vspace*{.5em} \noindent \thanks}
\newcommand{\beq}{\begin{equation}}
\newcommand{\eeq}{\end{equation}}
\newcommand{\Proof}{\begin{proof}}
\newcommand{\QED}{\end{proof} \noindent}

\newcommand{\la}{\langle}
\newcommand{\ra}{\rangle}
\newcommand{\bra}{\mathopen{\big<}}
\newcommand{\ket}{\mathclose{\big>}}
\newcommand{\C}{\mathbb{C}}
\newcommand{\R}{\mathbb{R}}
\newcommand{\1}{\mbox{\rm 1 \hspace{-1.05 em} 1}}

\newcommand{\N}{\mathbb{N}}
\renewcommand{\H}{\mathscr{H}}

\newcommand{\F}{{\mathscr{F}}}
\newcommand{\K}{{\mathcal{K}}}
\renewcommand{\P}{{\mathcal{P}}}
\newcommand{\M}{{\mathfrak{M}}}
\newcommand{\B}{{\mathfrak{B}}}
\renewcommand{\L}{{\mathcal{L}}}
\newcommand{\Sact}{{\mathcal{S}}}
\newcommand{\T}{{\mathcal{T}}}
\newcommand{\Lin}{\text{\rm{L}}}
\newcommand{\m}{{\mathfrak{m}}}
\newcommand{\n}{{\mathfrak{n}}}
\renewcommand{\O}{\mathscr{O}}

\DeclareMathOperator{\Tr}{Tr}

\DeclareMathOperator{\supp}{supp}

\DeclareMathOperator*{\esssup}{ess\,sup}

\begin{document}

\begin{abstract}
We derive the Euler-Lagrange equations for minimizers of causal variational principles in the non-compact
setting with constraints, possibly prescribing symmetries.
Considering first variations, we show that the minimizing measure is supported
on the intersection of a hyperplane with a level set of a function which is homogeneous of degree two.
Moreover, we perform second variations to obtain that the compact operator representing
the quadratic part of the action is positive semi-definite.
The key ingredient for the proof is a subtle adaptation of the Lagrange multiplier method
to variational principles on convex sets.
\end{abstract}

\maketitle
\tableofcontents

\section{Introduction} \label{sec11}
Causal variational principles arise in the context of relativistic quantum theory
(see~\cite{PFP, sector} or the review articles~\cite{rrev, srev}).
In~\cite{continuum} they were introduced in a broader
mathematical context, and the existence of minimizers was proved in various situations
(for previous existence results in the simpler discrete setting see~\cite{discrete}).
The structure of minimizers was first analyzed in~\cite{support} in the
compact setting without constraints. In the present paper, we turn attention to the general
non-compact setting involving constraints and possibly symmetries.
Analyzing first and second variations, we derive general results on the structure of minimizing measures.
Our results are important because they set up the mathematical framework and introduce the methods
needed for a detailed analysis of minimizers of causal variational principles.
Ultimately, our goal is to get a mathematical justification of the physical assumptions
on the vacuum minimizer as stated in~\cite[Chapters~4 and~5]{PFP} and~\cite[Section~3]{sector}
(for details on the regularization see~\cite{reg}).

Before delving into the main results, we briefly recall causal variational principles as introduced
in~\cite[Section~2]{continuum}, always specializing to the class of variational principles of interest here.
Let~$(M, \mu)$ be a measure space normalized by~$\mu(M)=1$.
For given integers~$k$ and~$n$ with~$k\geq 2n$, we let~$\F$ be the set of all Hermitian
$k \times k$-matrices of rank at most~$2n$, which (counting with multiplicities) have
at most~$n$ positive and at most~$n$ negative eigenvalues.
In a causal variational principle one
minimizes an action~$\Sact[F]$ under variations of a measurable function~$F : M \rightarrow \F$,
imposing suitable constraints. More specifically, for a given measurable function~$F : M \rightarrow \F$,
we let~$\rho = F_* \mu$ be the push-forward measure on~$\F$
(defined by~$\rho(\Omega) = \mu(F^{-1}(\Omega))$).
For any~$x, y \in \F$ we form the operator product
\beq \label{Adef}
A_{xy} = x \!\cdot\! y \::\: \C^k \rightarrow \C^k
\eeq
and denote its eigenvalues counted with algebraic multiplicities by
\beq \label{lambdacount}
\lambda_1^{xy}, \ldots, \lambda_{2n}^{xy}, 
\underbrace{0,\ldots, 0}_{\text{$k-2n$ times}}
\quad \text{with} \quad \lambda_j^{xy} \in \C \:.
\eeq
We define the {\em{spectral weight}}~$|A_{xy}|$ by
\beq \label{sweight1}
|A_{xy}| = \sum_{j=1}^{2n} |\lambda_j^{xy}| ,
\eeq
and similarly set~$|A_{xy}^2| = \sum_{j=1}^{2n} |\lambda_j^{xy}|^2$.
We introduce
\beq \label{Lagdef}
\text{the Lagrangian} \quad\quad \L[A_{xy}] = |A_{xy}^2| - \frac{1}{2n}\: |A_{xy}|^2
\eeq
and define the functionals~$\Sact$ and~$\T$ by
\begin{align}
\Sact &= \iint_{\F \times \F} \L[A_{xy}]\: d\rho(x)\, d\rho(y) \label{Sdef} \\
\T &= \iint_{\F \times \F} |A_{xy}|^2\: d\rho(x)\, d\rho(y) \label{Tdef}\:.
\end{align}
We also introduce the following constraints:
\begin{itemize}
\item[(BC)] The {\em{boundedness constraint}}: \qquad\;\;
$\T \leq C$ \\[-0.5em]
\item[(TC)] The {\em{trace constraint}}: \qquad\;\;
$\displaystyle \int_\F \Tr (x) \, d\rho(x) = k$ \\
\item[(IC)] The {\em{identity constraint}}: \qquad\;
$\displaystyle \int_\F x \, d\rho(x) = \1_{\C^k}\:.$
\end{itemize}
Our variational principle is to minimize~$\Sact$ by varying~$F$ in the class of all measurable
functions from~$M$ to~$\F$, under the constraints~(BC) and either~{\rm{(TC)}} or~{\rm{(IC)}}.
In~\cite[Theorem~2.3]{continuum} it is shown that the minimum of this variational
principle is attained by a function~$F \in L^2(M, \F, d\mu)$.

The measure space~$(M, \mu)$ may pose constraints on the form of the
push-forward measure~$\rho$ (for example, in the discrete setting one chooses~$\mu$ as the
normalized counting measure on~$M = \{1, \ldots, m \}$; then the support of~$\rho$ necessarily
consists of at most~$m$ points).
In what follows, we will always be concerned with the so-called {\em{continuous setting}}
where we do not want to impose any constraints on the form of the measure~$\rho$.
In technical terms, this can be achieved by assuming that the measure space~$(M, \mu)$ is non-atomic;
then the push-forward measure~$\rho$ can indeed be arranged to be any normalized positive
regular Borel measure on~$\F$ (see~\cite[Section~1.4 and Lemma~1.4]{continuum}).
This makes it possible to restrict attention to the measure~$\rho$ in the class
\beq \label{rhoclass}
\rho \in \M := \{ \text{normalized positive
regular Borel measures on~$\F$} \}\:,
\eeq
disregarding the measure space~$(M, \mu)$ and the function~$F$. This leads us to the
variational principles to be considered here:
\begin{Def} \label{defcausal}
For any parameter~$C > 0$, the {\bf{causal variational principle in the continuum setting}} is to
minimize~$\Sact$ by varying~$\rho \in \M$ under the constraints
\[ {\rm{(BC)}} \qquad \text{and either~{\rm{(TC)}} or~{\rm{(IC)}}}\:. \]
\end{Def} \noindent
Again, the existence of minimizers is proved in~\cite[Theorem~2.3]{continuum}.
The goal of this paper is to analyze the structure of a minimizing measure~$\rho$.

To clarify the terminology, we first remark that the spectral properties of~$A_{xy}$ induce
the following ``causal structure'' on the support of the measure~$\rho$
(for the connection to the physical notion of causality in space-time
we refer to~\cite{rrev, srev}).
\begin{Def} {\em{ Two points~$x, y \in \supp \rho \subset \F$ are
called {\em{timelike}} separated if the eigenvalues~$\lambda^{xy}_1, \ldots, \lambda^{xy}_n$ 
in~\eqref{lambdacount} are all real. They are said to be
{\em{spacelike}} separated if the~$\lambda^{xy}_j$ are all complex
and have the same absolute value. 
In all other cases, the points~$x$ and~$y$ are said to be {\em{lightlike}} separated. }}
\end{Def} \noindent
Our variational principle is ``causal'' in the sense that~$\L[A_{xy}]$ vanishes 
if~$x$ and~$y$ are spacelike separated.
Next, we point out that the set~$\F$ is a non-compact topological space;
this is what we mean by the {\em{non-compact setting}}. In contrast,
by prescribing the eigenvalues of the elements of~$\F$
(see the constraint~(C3) in~\cite[Section~2.1]{continuum}), one can arrange that~$\F$
is a compact manifold. This {\em{compact setting}} is analyzed in a more general
context in~\cite{support}. 
Unfortunately, for most of the methods used in~\cite{support} the compactness of~$\F$ is essential.
The present paper is the first analytic work on the structure of the minimizers of causal variational principles
in the non-compact setting.

The usual approach for treating variational principles with constraints is to apply the method of
Lagrange multipliers. For our variational principle, this method fails, essentially because positive measures
do not form a vector space (for details cf.\ Section~\ref{sec31} and Figure~\ref{figconstraint} below).
To circumvent this difficulty, in Section~\ref{sec3} we will develop an alternative method
which reproduces the results of Lagrange multipliers with subtle modifications.

Our main result can be understood heuristically from the standard Lagrange multiplier
method as follows. We add the constraints multiplied by Lagrange parameters~$\kappa, \Lambda, c$ to the
action so as to form the effective action
\beq \label{Seff}
\Sact_\text{eff} = \Sact + \kappa \T - \int_\F \Tr \big( \Lambda \!\cdot\! x \big)\: d\rho - c \int_\F d\rho \:,
\eeq
where in the case of the constraint~(TC), $\Lambda$ is a multiple of the identity matrix,
whereas in the case of~(IC), it can be any Hermitian $k \times k$-matrix.
The Lagrange multiplier~$c$ takes into account that~$\rho$ must be normalized.
Note that the positivity of the measure~$\rho$ cannot be encoded in terms of Lagrange multipliers.
Instead, we need to make sure in all our variations that~$\rho$ stays positive.
Considering for any~$x \in \F$ the first variation
\beq \label{var1}
\tilde{\rho}_\tau = \rho + \tau\, \delta_x \:,\qquad \tau \in [0, 1)
\eeq
(where~$\delta_x$ is the Dirac measure supported at~$x$; note that~$\tau$ is
non-negative in order to ensure that~$\tilde{\rho}_\tau$ is positive),
a short formal calculation yields the Euler-Lagrange inequality
\beq \label{EL1}
\Phi(x) - c \geq 0 \qquad \text{for all~$x \in \F$}\:,
\eeq
where
\beq \label{Phidef}
\Phi(x) := 2 \int_\F \big( \L(x,y) + \kappa\, |A_{xy}|^2 \big) \,d\rho(y) - \Tr(\Lambda \!\cdot\! x)\:.
\eeq
If the point~$x$ lies on the support of~$\rho$, we can extend the variation~\eqref{var1}
to small negative values of~$\tau$ (at least heuristically; to make the argument mathematically sound,
one needs to approximate the Dirac measure by a measure which is absolutely continuous
with respect to~$\rho$). When doing so, \eqref{EL1} becomes an equality,
\beq \label{EL2}
\Phi(x) - c = 0 \qquad \text{for all~$x \in \supp \rho$}\:.
\eeq
Combining~\eqref{EL1} with~\eqref{EL2}, we conclude that~$\Phi$ is minimal
on the support of~$\rho$. Accordingly,
\[ \frac{d}{dt} \Phi(t x)|_{t=1} = 0 \qquad \text{for all~$x \in \supp \rho$}\:. \]
This implies that the parts of~$\Phi$ which are homogeneous of degree two and one, denoted by
\begin{align}
\Phi_2(x) &:= 2 \int_\F \big( \L(x,y) + \kappa\, |A_{xy}|^2 \big) \,d\rho(y) \label{Phi2def} \\
\Phi_1(x) &:= \Phi(x) - \Phi_2(x) = - \Tr(\Lambda \!\cdot\! x) \:,
\end{align}
are related to each other by
\beq \label{EL3}
2 \Phi_2(x) + \Phi_1(x) = 0 \qquad \text{for all~$x \in \supp \rho$}\:.
\eeq
Now, combining~\eqref{EL2} and~\eqref{EL3} gives
\[ \Phi_1(x) = 2c = -2 \Phi_2(x) \:. \]
Integrating over~$x$, one can determine the constant~$c$.

\begin{figure}
\begin{center}
\begin{picture}(0,0)%
\includegraphics{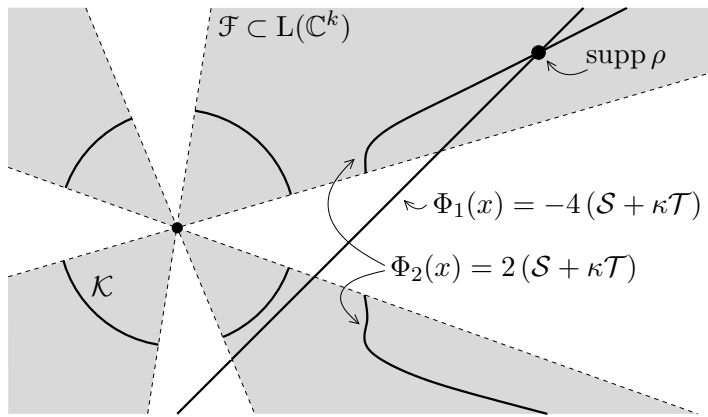}%
\end{picture}%
\setlength{\unitlength}{1865sp}%
\begingroup\makeatletter\ifx\SetFigFont\undefined%
\gdef\SetFigFont#1#2#3#4#5{%
  \reset@font\fontsize{#1}{#2pt}%
  \fontfamily{#3}\fontseries{#4}\fontshape{#5}%
  \selectfont}%
\fi\endgroup%
\begin{picture}(9471,5466)(4492,-6169)
\put(7294,-1102){\makebox(0,0)[lb]{\smash{{\SetFigFont{11}{13.2}{\rmdefault}{\mddefault}{\updefault}$\F \subset \Lin(\C^k)$}}}}
\put(5605,-4567){\makebox(0,0)[lb]{\smash{{\SetFigFont{11}{13.2}{\rmdefault}{\mddefault}{\updefault}${\mathcal{K}}$}}}}
\put(12187,-1466){\makebox(0,0)[lb]{\smash{{\SetFigFont{11}{13.2}{\rmdefault}{\mddefault}{\updefault}$\supp \rho$}}}}
\put(10141,-3459){\makebox(0,0)[lb]{\smash{{\SetFigFont{11}{13.2}{\rmdefault}{\mddefault}{\updefault}$\Phi_1(x)=-4\, (\Sact + \kappa \T)$}}}}
\put(9579,-4306){\makebox(0,0)[lb]{\smash{{\SetFigFont{11}{13.2}{\rmdefault}{\mddefault}{\updefault}$\Phi_2(x)=2\, (\Sact + \kappa \T)$}}}}
\end{picture}%
\caption{Example for the level sets of~$\Phi_1$ and~$\Phi_2$ and the support of~$\rho$.}
\label{figKf}
\end{center}
\end{figure}

The following theorem\footnote{For preliminary results and numerical
examples see the master thesis~\cite{eckl}, which also treats the case 
when the measure~$\rho$ is a counting measure. However, in this master thesis the
complication discussed in Figure~\ref{figconstraint} on page~\pageref{figconstraint} is disregarded.}
rigorously establishes this heuristic result under the additional assumption~\eqref{Cbound}.
\begin{Thm} \label{thmhyper} Suppose that~$\rho$ is a minimizer of the variational principle
of Definition~\ref{defcausal}, where the constant~$C$ satisfies the inequality
\beq \label{Cbound}
C > C_{\min} := \inf \big\{ \T(\mu) \:|\: \text{$\mu \in \M$ satisfies~(TC) respectively~(IC) } \big\} \:.
\eeq
Then for a suitable choice of the Lagrange multipliers
\[ \kappa \geq 0 \qquad \text{and} \qquad \Lambda \in \Lin(\C^k) \:, \]
the measure~$\rho$ is supported on the intersection of the level sets
\beq \label{hyper}
\Phi_1(x) = -4 \left( \Sact + \kappa \T \right)  \qquad \text{and} \qquad
\Phi_2(x) = 2 \left( \Sact + \kappa \T \right) \:.
\eeq
In the cases of the trace constraint~(TC) and the identity constraint~(IC),
the matrix~$\Lambda$ is a multiple of the identity and a general Hermitian matrix, respectively.
In the case~$\T(\rho) < C$, we may choose~$\kappa=0$.
\end{Thm} \noindent
This result is illustrated in Figure~\ref{figKf}. Note that the set~$\Phi_1^{-1}(-4 (\Sact + \kappa \T ))$
is a hyperplane in~$\Lin(\C^k)$. The set~$\Phi_2^{-1}(2 ( \Sact + \kappa \T ))$, on the other hand,
is the level set of a function which is homogeneous of degree two.
The support of~$\rho$ is contained in the intersection of these two sets.
This intersection might be non-compact. It is an open problem whether
the support of a minimizing measure is always compact.

The above theorem is supplemented by additional results, as we now briefly outline.
Theorem~\ref{thm2} gives sufficient conditions guaranteeing that the
function~$\Phi$ is indeed minimal on the support of~$\rho$.
When these conditions fail, a weaker statement can nonetheless
be obtained (Theorem~\ref{thm3}).
In Sections~\ref{sec34} and~\ref{sec35}, we consider second variations.
We prove that a suitable compact operator~$L$ on a Hilbert space is positive semi-definite
(Theorem~\ref{thm4}).
This positivity results bears similarity with~\cite[Lemma~4.5]{support} in the
compact setting.
In Theorem~\ref{thm5} we prove that the operator~$L$ stays positive when extended to
the direct sum of the aforementioned Hilbert space with any one-dimensional
vector space chosen within a specified class.
Section~\ref{sec36} is devoted to an a-priori estimate which shows in particular
that the support of~$\rho$ is compact
if the Lagrange multiplier~$\kappa$ is strictly positive.
Finally, in Section~\ref{secequi} we extend our results to a class of equivariant
variational principles.

\section{Preliminaries}
\subsection{Causal Fermion Systems}
We now briefly recall how the variational principles introduced in Definition~\ref{defcausal}
arise in the more general setting of causal fermion systems as introduced in~\cite[Section~1]{rrev}.
We first give the general definition.

\begin{Def} \label{defparticle} {\em{
Given a complex Hilbert space~$(\H, \la .|. \ra_\H)$ (the {\em{``particle space''}})
and a parameter~$n \in \N$ (the {\em{``spin dimension''}}), we let~$\F \subset \Lin(\H)$ be the set of all
self-adjoint operators on~$\H$ of finite rank, which (counting with multiplicities) have
at most~$n$ positive and at most~$n$ negative eigenvalues. On~$\F$ we are given
a positive measure~$\rho$ (defined on a $\sigma$-algebra of subsets of~$\F$), the so-called
{\em{universal measure}}. We refer to~$(\H, \F, \rho)$ as a {\em{causal fermion system in the
particle representation}}. }}
\end{Def} \noindent
Starting from this definition, one can construct a space-time endowed with a topological, causal
and metric structure, together with a collection of quantum mechanical wave functions in space-time
 (see~\cite{rrev} and~\cite{lqg}). We shall not enter these constructions here, but instead concentrate
 on the analytical aspects of the approach.

In order to get back to the setting of Section~\ref{sec11}, we specialize
the above framework in the following way: First, we assume that
particle space~$\H$ has finite dimension~$k$; then it can clearly be identified with the
Euclidean~$\C^k$. Moreover, we impose that~$\rho$ is in the class~\eqref{rhoclass}.
Then we can consider the variational principle of Definition~\ref{defcausal}.

In the case when~$\H$ is infinite-dimensional, the set~$\F \subset \Lin(\H)$ is
a topological space which is not locally compact. As a consequence,
causal variational principles are in general ill-defined (the physical picture is that
the limit~$\dim \H \rightarrow \infty$ corresponds to an idealized space-time where
the inherent ultraviolet regularization has been taken out).
However, if one assumes a symmetry group~$G$ which is so large that~$\F/G$ is locally compact,
then causal variational principles again make mathematical sense. This is
the equivariant setting which we will consider in Section~\ref{secequi}.

\subsection{The Moment Measures}\label{mommeas}
Let us assume that the measure~$\rho$ on~$\F$ is a minimizer of the variational principle
of Definition~\ref{defcausal}. We recall the definition of moment measures
as introduced in~\cite[Definition~2.10]{continuum}.
\begin{Def} \label{defmm}
Let~$\K$ be the compact topological space
\beq \label{Kdef}
\K = \{ p \in \F \text{ with } \|p\|=1 \} \cup \{0\} \:.
\eeq
We define the measurable sets of~$\K$ by the requirement that the sets
\[ \R^+ \Omega = \{ \lambda p \:|\: \lambda \in \R^+, p \in \Omega\} \]
and~$\R^- \Omega$ should be $\rho$-measurable in~$\F$. We introduce the measures~$\m^{(0)}$, 
$\m^{(1)}$ and~$\m^{(2)}$ by
\begin{align}
\m^{(0)}(\Omega) &= \frac{1}{2}\: \rho \big(\R_+ \Omega \setminus \{0\} \big) 
+ \frac{1}{2}\: \rho \big( \R_- \Omega \setminus \{0\} \big)
+ \rho \big( \Omega \cap \{0\} \big) \label{m0def} \\
\m^{(1)}(\Omega) &= \frac{1}{2} \int_{\R^+ \Omega} \|p\| \,d\rho(p) \:-\:
\frac{1}{2} \int_{\R^- \Omega} \|p\| \,d\rho(p) \label{m1def} \\
\m^{(2)}(\Omega) &= \frac{1}{2} \int_{\R_+ \Omega} \|p\|^2 \,d\rho(p) \:+\:
\frac{1}{2} \int_{\R_- \Omega} \|p\|^2 \,d\rho(p)
\:. \label{m2def}
\end{align}
The measure~$\m^{(l)}$ is referred to as the $l^\text{th}$ {\bf{moment measure}}.
\end{Def} \noindent

Exactly as in~\cite[Section~2.3]{continuum}, the homogeneity of our functionals yields that
\begin{align}
1 &= \rho(\F) = \m^{(0)}(\K) \label{zeromom} \\
\int_{\F} x\, d\rho(x) &= \int_{\K} x\, d\m^{(1)}(x)  \label{onemom} \\
\Sact(\rho) &= \iint_{\K \times \K} \L[A_{xy}]\: d\m^{(2)}(x)\, d\m^{(2)}(y) \label{twomom0} \\
\T(\rho) &= \iint_{\K \times \K} |A_{xy}|^2\: d\m^{(2)}(x)\, d\m^{(2)}(y) \:, \label{twomom}
\end{align}
making it possible to express the action as well as all the constraints in terms of the moment measures.
Moreover, the moment measures have the Radon-Nikodym decomposition
\[ d\m^{(1)} = f\, d\m^{(0)} \:,\qquad
d\m^{(2)} = |f|^2\, d\m^{(0)} + d\n\:, \]
where~$f \in L^2(\K, d\m^{(0)})$, and~$\n$ is a positive measure on~$\K$ which need not be
absolutely continuous with respect to~$\m^{(0)}$. If~$\n \neq 0$,
by setting~$\n$ to zero we can strictly decrease the action without violating our constraints
(see~\eqref{zeromom}--\eqref{twomom}). It follows that~$\n$ vanishes
for our minimizing measure~$\rho$. We thus obtain the representation of the moment measures
\beq \label{RN}
d\m^{(1)} = f\, d\m^{(0)} \:,\qquad
d\m^{(2)} = |f|^2\, d\m^{(0)}\:.
\eeq
From~\eqref{m1def} it is clear that~$f$ is odd,
\beq \label{fodd}
f(-x) = -f(x) \quad \text{for all~$x \in \K$}\:.
\eeq

The next proposition shows that the measure~$\rho$ is uniquely determined by the moment measures.
\begin{Prp}\label{momentprop} For a given normalized measure~$\m^{(0)}$ on~$\K$ and a given function~$f \in L^2(\K, d\m^{(0)})$ satisfying~\eqref{fodd}, there is a unique normalized measure~$\rho$ on~$\F$
such that the corresponding moment measures~\eqref{m0def}-\eqref{m2def} have the
Radon-Nikodym representation~\eqref{RN}. The measure~$\rho$ is supported on the
graph of~$f$ over~$\K$, i.e.\
\beq \label{graph}
\supp \rho \subset \overline{ \left\{ f(x)\, x \;\text{ with }\; x \in \K \right\} }\: .
\eeq
\end{Prp}
\Proof The construction of the measure~$\rho$ is inspired by~\cite[Lemma~2.14]{continuum}.
A subset~$\Omega \subset \F$ is called $\rho$-measurable if the function~$\chi_\Omega \big( f(x)\, x \big)$
is~$\m^{(0)}$-measurable on~$\K$ (where~$\chi_\Omega$ denotes the characteristic function).
On the $\rho$-measurable sets we define the measure~$\rho$ by
\beq \label{rhorep}
\rho(\Omega) = \int_{\K} \chi_\Omega \big( f(x)\, x \big)\, d\m^{(0)}(x) \:.
\eeq
Obviously, the measure~$\rho$ is normalized and has the support property~\eqref{graph}.
Moreover, it is straightforward to verify that
for all~$l>0$,
\[ \int_{\R^+ \Omega} \|p\|^l\: d\rho = \int_\Omega |f(x)|^l\:\chi_{\{f > 0\}}(x)\: d\m^{(0)}(x) \:. \]
Using this identity, a direct computation shows that the moment measures corresponding to~$\rho$
indeed satisfy~\eqref{RN}.

To prove uniqueness, suppose that~$\rho$ is a measure with moment measures satisfying~\eqref{RN}.
Then for every $\m^{(0)}$-measurable set~$\Omega$,
\begin{align}
\frac{1}{2} &\int_{\R^+ \Omega} \!\big( \|p\| - f(p) \big)^2\: d\rho
+ \frac{1}{2} \int_{\R^- \Omega} \!\big(-\|p\| - f(p) \big)^2\: d\rho
\;+\; f(0)^2\: \m^{(0)}\big( \Omega \cap \{0\} \big) \label{first} \\
&= \m^{(2)}(\Omega) - 2 \int_\Omega f\, d\m^{(1)} + \int_\Omega f^2\, d\m^{(0)} = 0\:,
\end{align}
where we multiplied out and used~\eqref{RN}. In particular, both
integrands in~\eqref{first} must vanish almost everywhere.
Now a short calculation yields that~$\rho$ coincides with the measure~\eqref{rhorep}.
\QED
In order to clarify the meaning of~\eqref{graph}, we note that~$f \in L^2(\K, d\m^{(0)})$
stands for an equivalence class of functions which differ on a set of measure zero.
The right hand side of~\eqref{graph} may depend on the choice of the representative.
The above proposition states that the inclusion~\eqref{graph} holds for {\em{any}} choice of
the function~$f \in L^2(\K, d\m^{(0)})$.

\section{The Euler-Lagrange Equations} \label{sec3}

\subsection{Treating the Constraints} \label{sec31}
Considering on the set~$\F \subset \Lin(\C^k)$ the topology induced by
the sup-norm~$\| . \|$ on~$\Lin(\C^k)$, this set is a locally compact topological space.
Its subset~$\K \subset \F$ defined by~\eqref{Kdef} is compact.
Let~$\mu$ be a regular, locally finite Borel measure on~$\F$ (which is real, but not necessarily
positive; such measures are also called signed Radon measures).
Moreover, we assume that the following integral is finite,
\beq \label{2mom}
\|\mu\|_\B := \int_{\F} \big(1+\|x\|^2 )\, d|\mu|(x) < \infty
\eeq
(here~$| \mu |$ denotes the total variation of the measure~$\mu$; see for example~\cite[Section~6.1]{rudin}).
We denote the vector space of such measures by~$\B$.

\begin{Lemma} $(\B, \| . \|_\B)$ is a Banach space.
\end{Lemma}
\Proof It is obvious that $\|.\|_{\B}$ satisfies the axioms of a norm.
Thus it remains to show that this norm is complete. We first note that
\beq
\|\mu\|_\B \geq |\mu|(\F)\:.
\eeq
Accordingly, if $(\mu_j)_{j \in \N}$ is a Cauchy sequence in the norm $\| . \|_\B$,
then for every~$\eta \in C^0_0(\F, \R)$, the sequence of real numbers~$(|\mu_j|(\eta))_{j\in\mathbb{N}}$ is a
Cauchy sequence. A classical result on Radon measures (see for example~\cite[eq.~(13.4.1)]{dieudonne2}) guarantees that the sequence~$(\mu_j)$ converges as Radon measures to some limit
measure~$\mu$. It remains to show that the limit measure satisfies the condition~\eqref{2mom}.
We already know from the above argument that
\beq\label{convmuj}
\lim_{j\rightarrow\infty}\,\mu_j(\eta)\;=\;\mu(\eta)\qquad\forall\:\eta \in C^0_0(\F,\mathbb{R})\:.
\eeq
We next fix $r>1$, and let $\eta_r:[0,\infty)\rightarrow[0,1]$ be a continuous cut-off function satisfying
\beq
\eta_r(t)\;=\;\left\{\begin{array}{lll}1&\;\;\;& \text{if~$t\leq r$} \\[1ex] 0&& \text{if~$t> r+r^{-1}$} \:.\end{array}\right.
\eeq
Then the function
\beq
x\in\F\;\longmapsto\;\big( 1+\|x\|^2 \big) \:\eta_r\big(\|x\|\big)
\eeq
is continuous with compact support in~$B_{r+r^{-1}}$, where~$B_r$ denotes the
open ball in~$\F$,
\beq
B_r := \big\{x\in\F \text{ with } \|x\| < r \big\} \subset \F \:.
\eeq
Whence, from (\ref{convmuj}), there holds
\beq
\lim_{j\rightarrow\infty}\int_{\F} \big( 1+\|x\|^2 \big) \:\eta_r\big(\|x\|\big)\,d|\mu_j|(x)
=\int_{\F}\big( 1+\|x\|^2 \big) \:\eta_r\big(\|x\|\big)\,d|\mu|(x)\:.
\eeq
It follows accordingly that
\begin{align*}
\int_{B_r} \big( &1+\|x\|^2 \big)\, d|\mu|(x) \leq \int_{\F} \big( 1+\|x\|^2 \big) \,\eta_r\big(\|x\|\big)\,d|\mu|(x) \\
&= \lim_{j\rightarrow\infty}\,\int_{\F} \big( 1+\|x\|^2 \big) \,\eta_r\big(\|x\|\big)\,d|\mu_j|(x)
\leq \lim_{j\rightarrow\infty}\,\|\mu_j\|_\B \:,
\end{align*}
and the last limit is bounded uniformly in~$r>1$. As $\F$ is locally compact, on the left hand side we may
pass to the limit~$r\nearrow\infty$ to obtain that~$\|\mu\|_\B$ is finite. This concludes the proof.
\QED

The definitions~\eqref{Sdef} and~\eqref{Tdef} of the functionals~$\Sact$ and~$\T$
as well as the definition of the moment measures (see Definition~\ref{defmm}) can be
extended in a straightforward way to a real measure~$\rho \in \B$.
We now estimate these objects in terms of the norm~$\| . \|_\B$.
\begin{Prp} There is a constant~$c=c(\F)>0$ such that
\begin{align}
|\Sact(\mu)|,\, |\T(\mu)| &\leq c\, \|\mu\|^2_\B  &&\hspace*{-2cm} \text{for all~$\mu \in \B$} \label{STup} \\
\|\rho\|^2_\B &\leq 2 + c\, \T(\rho) &&\hspace*{-2cm} \text{for all~$\rho \in \M$} \:. \label{STlow}
\end{align}
\end{Prp}
\Proof Estimating the integrals in Definition~\ref{defmm} by~\eqref{2mom}, one readily finds that
\beq \label{momes}
|\m^{(0)}|(\K),\, |\m^{(1)}|(\K),\, |\m^{(2)}|(\K) \leq \|\rho\|_\B \qquad \text{for all~$\rho \in \B$}\:.
\eeq

The functions~$\L$ and~$|A_{xy}|^2$ are clearly continuous on~$\K \times \K$.
As~$\K$ is compact, they are bounded,
\[ \L(x,y), |A_{xy}|^2 \leq c \qquad \text{for all~$x, y \in \K$}\:. \]
Using these inequalities in~\eqref{twomom0} and~\eqref{twomom},
we can apply~\eqref{momes} to obtain~\eqref{STup}.

In order to derive~\eqref{STlow}, we first note that since every measure~$\rho \in \M$ is normalized and positive,
\[ \| \rho \|_\B = \m^{(0)}(\K) + \m^{(2)}(\K) = 1 + \m^{(2)}(\K) \:. \]
Now we can apply the lower bound on~$\m^{(2)}(\K)$ in~\cite[Lemma~2.12]{continuum}.
\QED

The inequality~\eqref{STlow} implies that a minimizer~$\rho \in \M$ of our variational principle
will be a vector in~$\B$. This makes it possible to consider our variational principle
on the subset~$\M \cap \B$ of the Banach space~$\B$.
Usually, constraints of variational principles are treated with Lagrange multipliers.
We now explain why this method cannot be applied in our setting.
Our first constraint is that we vary in the subset of positive measures. This corresponds
to an infinite number of inequality constraints (namely~$\rho(\Omega) \geq 0$ for all
measurable~$\Omega \subset \F$), making it impossible to apply standard Lagrange multipliers.
The normalization of~$\rho$ could be treated as in~\eqref{Seff} by a Lagrange multiplier.
But as the normalization of~$\rho$ can always be arranged by rescaling, there is no advantage
in doing so. Instead, it is preferable to consider
the minimization problem on the convex subset~$\M \cap \B$ of the Banach space~$\B$.

\begin{figure}
\begin{picture}(0,0)%
\includegraphics{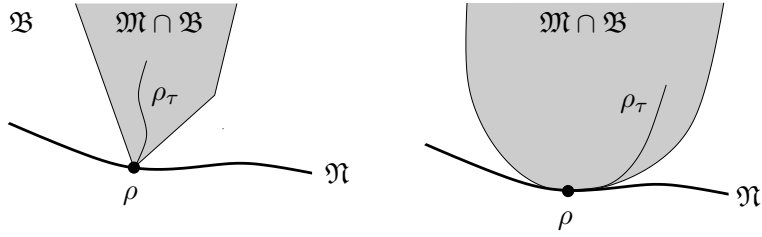}%
\end{picture}%
\setlength{\unitlength}{1533sp}%
\begingroup\makeatletter\ifx\SetFigFont\undefined%
\gdef\SetFigFont#1#2#3#4#5{%
  \reset@font\fontsize{#1}{#2pt}%
  \fontfamily{#3}\fontseries{#4}\fontshape{#5}%
  \selectfont}%
\fi\endgroup%
\begin{picture}(11823,3740)(47,-2867)
\put(1936,-2326){\makebox(0,0)[lb]{\smash{{\SetFigFont{11}{13.2}{\rmdefault}{\mddefault}{\updefault}$\rho$}}}}
\put(2396,-716){\makebox(0,0)[lb]{\smash{{\SetFigFont{11}{13.2}{\rmdefault}{\mddefault}{\updefault}$\rho_\tau$}}}}
\put(8971,-2731){\makebox(0,0)[lb]{\smash{{\SetFigFont{11}{13.2}{\rmdefault}{\mddefault}{\updefault}$\rho$}}}}
\put(9978,-840){\makebox(0,0)[lb]{\smash{{\SetFigFont{11}{13.2}{\rmdefault}{\mddefault}{\updefault}$\rho_\tau$}}}}
\put(5199,-2063){\makebox(0,0)[lb]{\smash{{\SetFigFont{11}{13.2}{\rmdefault}{\mddefault}{\updefault}${\mathfrak{N}}$}}}}
\put(11855,-2430){\makebox(0,0)[lb]{\smash{{\SetFigFont{11}{13.2}{\rmdefault}{\mddefault}{\updefault}${\mathfrak{N}}$}}}}
\put( 91,374){\makebox(0,0)[lb]{\smash{{\SetFigFont{11}{13.2}{\rmdefault}{\mddefault}{\updefault}${\mathfrak{B}}$}}}}
\put(1823,374){\makebox(0,0)[lb]{\smash{{\SetFigFont{11}{13.2}{\rmdefault}{\mddefault}{\updefault}${\mathfrak{M}} \cap {\mathfrak{B}}$}}}}
\put(8676,374){\makebox(0,0)[lb]{\smash{{\SetFigFont{11}{13.2}{\rmdefault}{\mddefault}{\updefault}${\mathfrak{M}} \cap {\mathfrak{B}}$}}}}
\end{picture}%
\caption{Minimizing in the convex subset~$\M \cap \B$ with constraints.}
\label{figconstraint}
\end{figure}

We would like to treat the constraint~(BC) as well as the additional constraints~(TC) or~(IC)
with Lagrange multipliers. The fact that~(BC) is an inequality constraint does not cause difficulties,
because for variations which decrease~$\T$, we can disregard this constraint, whereas
for variations which increase~$\T$ we can impose the equality constraint~$\T=C$.
However, a general problem arises from the fact that we minimize only in a convex
subset~$\M \cap \B \subset \B$. The basic difficulty is seen most easily in the examples
shown in Figure~\ref{figconstraint}.
Assume for simplicity that we only have equality constraints and that we
are in the regular setting where the measures which satisfy the
constraints form a smooth Banach submanifold~$\mathfrak{N} \subset \B$.
Then~$\mathfrak{N}$ can be described locally as the zero set of a
function
\beq G : \B \rightarrow \R^L \:. \label{Gdef}
\eeq
The standard multiplier method would give parameters~$\lambda_l \in \R$ such that
\beq \label{lagrequal}
\frac{d}{d\tau} \Big( \Sact(\rho_\tau) - \sum_{l=1}^L \lambda_l \,G_l(\rho_\tau) \Big) \Big|_{\tau=0} = 0
\eeq
for any variation~$(\rho_\tau)_{\tau \geq 0}$.
Since we are only allowed to vary in the convex subset~$\M \cap \B$, it may happen that the
minimum is attained on the boundary of~$\M \cap \B$. In this case, we cannot expect
that equality holds in~\eqref{lagrequal}. Instead, one might expect naively the corresponding inequality
\beq \label{lagrinequ}
\frac{d}{d\tau} \Big( \Sact(\rho_\tau) - \sum_{l=1}^L \lambda_l \,G_l(\rho_\tau) \Big) \Big|_{\tau=0} \geq 0 \:,
\eeq
which should hold for any variation~$(\rho_\tau)_{\tau \in [0,1)}$ in~$\B \cap \M$.
However, this naive guess is not correct, as is illustrated in Figure~\ref{figconstraint}.
In the example on the left, the convex set~$\M \cap \B$
intersects~$\mathfrak{N}$ only in one point~$\rho$. Then~$\rho$ is clearly a minimizer
in~$\M \cap \B$ subject to the constraints, simply because there are no non-trivial variations of~$\rho$.
But the minimizing property does not give us any information on the variation~$\rho_\tau \in \B$.
In particular, there is no reason why~\eqref{lagrinequ} should hold.
In the example on the right of Figure~\ref{figconstraint}, $\rho$ is again a trivial minimizer
in~$\M \cap \B$ subject to the constraints. There is even a variation~$(\rho_\tau)_{\tau \in [0,1)}$
in~$\M \cap \B$ which is tangential to~$\mathfrak{N}$, implying that the resulting Lagrange multiplier
terms in~\eqref{lagrinequ} vanish. Choosing the action such
that~$\partial_\tau \Sact(\rho_\tau)|_{\tau=0}<0$, one can construct examples where~\eqref{lagrinequ}
is violated.

Our method to overcome this difficulty is to first derive
an inequality which shows that that for any variation~$\rho_\tau \in \M \cap \B$, the first variation of the action
is bounded from below by the first variation of the constraint functions
(see Proposition~\ref{prplagrange} below).
This result is much weaker than the inequality~\eqref{lagrinequ}, 
basically because the Lagrange multiplier terms are replaced by an estimate of their absolute values.
Despite this rough estimate, Proposition~\eqref{prplagrange} will be very useful for analyzing
the minimizing measure. More precisely, in Section~\ref{sec32} we shall apply it to 
special variations~$\rho_\tau$ for which~$\partial_\tau G(\rho_\tau)|_{\tau=0}$ vanishes.
Then the error term in~\eqref{Seffp} drops out, giving a sharp inequality.
Before stating our result, we need to specify the functions
which describe the constraints. The constraints~(TC) and~(IC) are linear in the measure;
we denote their total number by~$L$.
For the constraint~(TC), we choose~$L=1$ and
\beq \label{G0}
G_1(\mu) = k - \int_\F \Tr(x)\: d\mu(x)\:.
\eeq
For the constraint~(IC), we set~$L=k (k+1)/2$.
Choosing a basis~$e_1, \ldots, e_L$ of the symmetric~$k \times k$-matrices,
we let
\beq \label{G1}
G_l(\mu) = \Tr \bigg( e_l \Big( \1_{\C^k} - \int_F x\: d\mu(x) \Big) \bigg)\:,\qquad l=1,\ldots, L \:.
\eeq
It is convenient to choose~$e_1=\1$, so that~\eqref{G0} agrees with~\eqref{G1} for~$l=1$.
Moreover, it is convenient to choose the matrices~$e_2, \ldots, e_L$ to be trace-free.

\begin{Prp} \label{prplagrange} Assume that~$\rho$ is a minimizer of the 
variational principle of Definition~\ref{defcausal}, where the constant~$C$ satisfies~\eqref{Cbound}.
Then there is a constant~$c$ such that 
for every $\B$-Fr{\'e}chet differentiable
family of measures $(\rho_\tau)_{\tau \in [0, 1)}$
in~$\B \cap \M$ with~$\rho_0=\rho$, the first variation satisfies the inequality
\beq \label{Seffp}
\begin{split}
\frac{d}{d\tau} \Sact(\rho_\tau) \Big|_{\tau=0} \geq &- c \; \bigg\| \frac{d}{d\tau} G(\rho_\tau)
\Big|_{\tau=0}\, \bigg\|_{\R^L} \\[0.2em]
&- \left\{ \begin{array}{ll} 0 & \text{if~$\T(\rho)<C$} \\[0.5em]
\displaystyle c \,\max \left( 0, \frac{d}{d\tau} \T(\rho_\tau) \Big|_{\tau=0} \right)
& \text{if~$\T(\rho)=C$} \:. \end{array} \right. 
\end{split}
\eeq
\end{Prp} \noindent
The method of the proof is to construct a corresponding
variation~$\tilde{\rho}_\tau \in \M \cap \B$ which also satisfies all the constraints
and then to exploit the inequality~$\partial_\tau \Sact(\tilde{\rho}_\tau)|_{\tau=0} \geq 0$.
In this construction, the assumption~\eqref{Cbound} will be used to rule out
degenerate cases as discussed in Figure~\ref{figconstraint}.
Unfortunately, it is impossible to write the difference
of the first variations~$\partial_\tau (\Sact(\rho_\tau)-\Sact(\tilde{\rho}_\tau))$
as a derivative of the constraints.

The proof of Proposition~\ref{prplagrange} is split up into several lemmas; it will
be completed towards the end of this section.
\begin{Lemma} \label{lemmafrechet}
The functions~$\Sact$, $\T$ and~$G$ are continuously Fr{\'e}chet differentiable.
\end{Lemma}
\Proof The inequality~\eqref{STup} implies that~$\Sact$ and~$\T$
are bounded bilinear functionals on~$\B \times \B$. Thus they are Fr{\'e}chet differentiable
at any~$\mu \in \B$ and
\begin{align}
(D \Sact)_\mu(\nu) &= 2 \iint_{\F \times \F} \L[A_{xy}]\,d\mu(x)\,d\nu(y) \label{diffS} \\
(D \T)_\mu(\nu) &= 2\iint_{\F \times \F} |A_{xy}|^2\,d\mu(x)\,d\nu(y) \:. \label{diffT}
\end{align}
More precisely, $D \Sact_\mu \in \B^*$ and
\[ \|D \Sact_\mu\|_{\B^*} := \sup_{\nu \in \B,\, \|\nu\|_\B=1} \big| (D \Sact)_\mu[\nu] \big| 
\:\leq\: c\, \|\mu\|_\B \:, \]
where in the last step we used~\eqref{STup}.
As the functionals~\eqref{diffS} and~\eqref{diffT} clearly depend continuously on~$\mu$,
we conclude that~$\Sact$ and~$\T$ are indeed in~$C^1(\B)$.
It remains to consider the functions~\eqref{G0}
and~\eqref{G1}. These are linear in~$\mu$, and the estimate
\[ \int_\F \,\Vert x\Vert\,d|\mu|(x) \leq \int_\F \,(1 + \Vert x\Vert^2)\,d|\mu|(x)
= \Vert\mu\Vert_\B \qquad\forall\:\mu\in\B \]
readily shows that their derivative is a bounded linear functional.
As this functional is continuous in~$\mu$ (it is even independent of~$\mu$),
it follows that~$G \in C^1(\B)$.
\QED

In the next lemma we construct measures for prescribed linear constraints
but such that the value of~$\T$ is smaller than that of a given minimizer.
For the construction we rescale the argument of a measure. We denote this
operation by~$\mathfrak{s}$,
\beq \label{scaledef}
\mathfrak{s}\::\: \R \times \B \rightarrow \B \:,\quad (\mathfrak{s}_\tau \mu)(\Omega) := \mu(\tau \Omega)\:.
\eeq
Obviously, $\mathfrak{s}_\tau$ maps~$\M \cap \B$ to itself.
\begin{Lemma} \label{lemmarhoh} For a given minimizer~$\rho \in \M \cap \B$, there is a parameter~$\delta>0$
and a smooth mapping $\hat{\rho} \::\: B_\delta(\rho) \subset \B \rightarrow \M \cap \B$
such that for all~$\mu \in B_\delta(\rho)$,
\beq \label{Trho}
G\big( \mu-\hat{\rho}(\mu) \big) = 0 \qquad \text{and} \qquad
\T(\hat{\rho}(\mu)) < C\:.
\eeq
Moreover, the measure~$\hat{\rho}$ satisfies the inequality
\beq \label{DTrho}
D \T|_\mu \,\hat{\rho} < 2 C \:.
\eeq
\end{Lemma}
\Proof According to the assumption~\eqref{Cbound}, there is a measure~$\rho_1 \in \M \cap \B$ such that
\[ \int_\F x\, d\rho_1 = \1_{\C^k} \qquad \text{and} \qquad \T(\rho_1) < C\:. \]
In the case of the identity constraint~(IC), we choose
additional measures~$\rho_2, \ldots, \rho_L \in \M \cap \B$ such that the matrices
\beq \label{linmat}
\int_\F x\, d\rho_l \:, \qquad l=1,\ldots, L
\eeq
are linearly independent (for example, these measures can be chosen as Dirac measures
supported at certain~$x \in\F$).

For parameters~$\kappa \in (0, L^{-1})$ and~$\tau \in \R^L$, we consider the family of measures
\[ \hat{\rho}(\kappa, \tau_1, \ldots, \tau_L) =
(1-\kappa L) \, \mathfrak{s}_{(1-\kappa L)^{-1}} \rho_1 + \kappa \sum_{l=1}^L
\mathfrak{s}_{\tau_l} \,\rho_l \:. \]
Then the functional~$G$ depends linearly on the parameters~$\tau_1, \ldots, \tau_L$,
and the mapping~$(\tau_1, \ldots, \tau_L) \mapsto (G_1, \ldots, G_L)$ is invertible.
Moreover, by choosing the parameters~$\kappa$ and~$\tau_l$ sufficiently small,
we can arrange by continuity that~$\T(\hat{\rho})<C$.
Finally, a direct computation shows that the measure~$\hat{\rho}$
is positive and normalized.

By continuity, it suffices to derive~\eqref{DTrho} for~$\mu = \rho$.
To this end, we consider the family of measures
\beq \label{tilderho}
\tilde{\rho}_\tau = \tau \hat{\rho} + (1-\tau)\, \rho \:.
\eeq
Then in view of~\eqref{TG} and~\eqref{diffT},
\[ \T(\rho_\tau) = \tau^2\, \T(\hat{\rho}) + \tau\, (1-\tau)\: D \T|_{\rho}\, \hat{\rho}
+ (1-\tau)^2\: \T(\rho)\:. \]
This functional is obviously quadratic in~$\tau$, and as~$\lim_{\tau \rightarrow \pm \infty} \T(\rho_\tau)=\infty$,
it is convex. Hence
\[ D \T|_{\rho}\, \hat{\rho} - 2 \T(\rho) = \frac{d}{d\tau} \T(\rho_\tau) \big|_{\tau=0} \leq
\T(\hat{\rho}) - \T(\rho) \]
and thus
\[ D \T|_{\rho}\, \hat{\rho} \leq \T(\hat{\rho}) + \T(\rho) \:. \]
Since~$\T(\hat{\rho}) < C$ and~$\T(\rho) \leq C$, we obtain the strict inequality~\eqref{DTrho}.
\QED

\begin{Lemma} \label{lemmaad}
Under the assumptions of Proposition~\ref{prplagrange},
for every minimizer~$\rho \in \M \cap \B$ there
are parameters~$\varepsilon, \delta>0$ and a continuous mapping
\[ \Phi \::\: (B_\delta(\rho) \subset \B) \times (B_\varepsilon(0) \subset \R^L) 
\times [0, \varepsilon) \rightarrow \B \]
with the following properties:
\begin{itemize}
\item[(a)] $\Phi(\mu,0,0) = \mu$ for all~$\mu \in B_\delta(\rho)$.
\item[(b)] For every~$t \in B_\varepsilon(0)$ and~$\tau \in [0, \varepsilon)$,
the function~$\Phi(., t, \tau) : B_\delta(\rho) \rightarrow \B$ maps the set $\M \cap B_\delta(\rho)$ to itself.
\item[(c)] The composition~$G \circ \Phi$ is in~$C^1(B_\delta(\rho) \times B_\varepsilon(0) \times[0, \varepsilon),
\R^L)$. Moreover, the~$L \times L$-matrix $D_2 (G \circ \Phi)|_{(\rho, 0,0)}$ is invertible
and~$D_3 (G \circ \Phi)|_{(\rho, 0, 0)} = 0$.
\item[(d)] The directional derivatives~$u \cdot D_2 (\T \circ \Phi)|_{(\rho, 0, 0)}$ (with~$u \in \R^L$)
and the partial derivative~$D_3 (\T \circ \Phi)|_{(\rho,0,0)}$ exist. They satisfy the inequalities
\begin{align*}
|u \cdot D_2 (\T \circ \Phi)(\rho, 0, 0)| &\leq c \, \|u\|_{\R^L} \\
D_3 (\T \circ \Phi)|_{(\rho,0,0)} &< 2 \,(C-\T(\rho))
\end{align*}
with a constant~$c=c(\rho)$.
\end{itemize}
\end{Lemma}
\Proof We make the ansatz
\begin{align}
\Phi(\mu, t, \tau)  &= (1-\tau) \:\sigma \,\mathfrak{s}_{\sigma^{-1}}\,\mu
+ \tau\, \hat{\rho} \label{c1} \\
&\quad+ \sum_{l=1}^L \Big( \max(t_l, 0)\: \rho_l
+ \max(-t_l, 0)\: \mathfrak{s}_{-1} \,\rho_l \Big) \:, \label{c2}
\end{align}
where
\[ \sigma =  \frac{1}{1-\tau} \Big( 1- \sum_{l=1}^L |t_l| - \tau \Big) \:. \]
Obviously, $\Phi$ is trivial in the case~$t=0$ and~$\tau=0$,
giving property~(a). Moreover, if~$t$ and~$\tau$ are sufficiently small,
we have a convex combination of measures, proving property~(b).

We point out that~$\Phi$ is {\em{not}} differentiable in~$t$
because of the absolute values and the factors~$\max(\pm t_l, 0)$.
On the other hand, this property is not needed, as we only claim that~$G \circ \Phi$
is differentiable. Lemma~\ref{lemmarhoh} yields that~$D_3 (G \circ \Phi)|_{(\rho, 0, 0)} = 0$.
But the linear constraints depend on the parameters~$t_l$.
Our ansatz ensures that this dependence is smooth even if some of the
parameters~$t_l$ vanish (note that~$G(\sigma \,\mathfrak{s}_{\sigma^{-1}} \mu)$ is independent of~$\sigma$
by homegeneity). Finally, as the matrices~\eqref{linmat} are linearly independent,
it follows immediately that~$D_2 (G \circ \Phi)|_{(\mu, t, \tau)}$ has maximal rank.
This proves~(c).

In order to prove~(d), we consider the functional~$\T \circ \Phi$.
Note that, again due to the absolute values and the factors~$\max(\pm t_l, 0)$, this functional is not differentiable
in the parameters~$t_l$. But clearly, the directional derivatives at~$t=0$ exist and are bounded.
Finally, the derivative with respect to~$\tau$ is computed with the help of~\eqref{DTrho}.
\QED

\Proof[Proof of Proposition~\ref{prplagrange}] Let us apply Lemma~\ref{lemmaad}.
First, as~$G \circ \Phi$ is continuously differentiable, we can conclude from~(c)
that there is~$\delta>0$ such that
the matrix~$D_2 \big(G \circ \Phi \big) (\mu, 0,\tau)$ is invertible for all~$\mu \in B_\delta(\rho) \subset \B$
and all~$\tau \in [0, \varepsilon)$.
Thus~$(G \circ \Phi)(\mu, ., \tau)$ is a local diffeomorphism, implying that (possibly after decreasing~$\delta$)
there is a mapping~$h \in C^1(B_\delta(\rho) \times [0, \varepsilon), B_\varepsilon(0))$ such
that~$h(\rho,0)=0$ and
\beq \label{Gsat}
(G \circ \Phi)(\mu, h(\mu, \sigma), \sigma) = 0 \qquad \text{for all~$\mu \in B_\delta(\rho)$
and~$\sigma \in [0, \varepsilon)$}\:.
\eeq

Let~$(\rho_\tau)_{\tau \in [0, 1)}$ be a variation in~$\B \cap \M$ with~$\rho_0=\rho$.
We choose~$\sigma = \kappa \tau$ with a constant~$\kappa>0$ to be determined later.
Then, using that~$h(\rho,0)=0$ and that~$D_3(G \circ \Phi)_{(\rho,0,0)}=0$, we obtain
\begin{align}
0 &= \frac{d}{d\tau} (G \circ \Phi)(\rho_\tau, h(\rho_\tau, \kappa \tau), \kappa \tau) \Big|_{\tau=0} \nonumber \\
&= D_1(G \circ \Phi)|_{(\rho, 0,0)} \,\dot{\rho}_0 
+ D_2(G \circ \Phi)|_{(\rho, 0,0)} \circ Dh|_{(\rho,0)} \,\dot{\rho}_0 \:. \label{GPhi}
\end{align}
We now introduce for~$\tau \in [0, \alpha)$ and sufficiently small~$\alpha>0$ the variation
\beq \label{trho}
\tilde{\rho}_\tau = \Phi \big( \rho_\tau, h(\rho_\tau, \kappa \tau), \kappa \tau \big) \:.
\eeq
In view of~(b) and~\eqref{Gsat}, this variation lies in~$\M \cap \B$ and satisfies the linear constraints.
Moreover, by choosing~$K$ sufficiently large, we can arrange in view of~(d) that
this variation decreases~$\T$. Thus it satisfies all the constraints and is
admissible for our variational principle. The minimality of~$\rho$ implies that
\[ 0 \leq \frac{d}{d\tau} \Sact(\tilde{\rho}_\tau) \Big|_{\tau=0} 
= D \Sact|_\rho\, \frac{d}{d\tau} \Phi(\rho_\tau, h(\rho_\tau, \kappa \tau), \kappa \tau) \Big|_{\tau=0} \:. \]
Computing the one-sided derivatives with the chain rule, we obtain
\[ \frac{d}{d\tau} \Phi(\rho_\tau, h(\rho_\tau)) \Big|_{\tau=0^+}
= \dot{\rho}_0 + E \:, \]
where the error term is bounded by
\[ \| E \| \leq c\, \big\| Dh|_{(\rho,0)} \,(\dot{\rho}_0, \kappa) \big\| + c\: \kappa \:. \]

In the case~$\T(\rho)<C$, we can choose~$\kappa=0$. 
Differentiating~\eqref{Gsat}, we obtain
\[ 0 = \frac{d}{d\tau} (G \circ \Phi)|_{(\rho_\tau, h(\rho_\tau,0), 0)}
= DG|_\rho \dot{\rho}_0 + D_2(G \circ \Phi)|_{(\rho,0,0)} \,Dh|_{(\rho,0)} (\dot{\rho}_0, 0) \:, \]
showing that~$Dh$ can be estimated in terms of the first derivatives of~$G$. This gives the result.

In the case~$\T(\rho)=C$, we know from~(d) that~$D_3(\T \circ \Phi)|_{(\rho,0,0)} < 0$.
Thus by choosing~$\kappa$ sufficiently large, we can compensate the
positive contribution to the variation of~$\T$ caused by~$\rho_\tau$
and by~$h$. Clearly, the parameter~$\kappa$ is bounded in terms of the variation of~$G$
and the positive part of~$\partial_\tau \T(\rho_\tau)|_{\tau=0}$. This concludes the proof.
\QED

We finally show how Proposition~\ref{prplagrange} can be adapted to second variations.
\begin{Prp} \label{prplagrange2} Assume that~$\rho$ is a minimizer of the 
variational principle of Definition~\ref{defcausal}, where the constant~$C$ satisfies~\eqref{Cbound}.
Then there is a constant~$c$ such that 
for every twice $\B$-Fr{\'e}chet differentiable
family of measures $(\rho_\tau)_{\tau \in [0, 1)}$
in~$\B \cap \M$ with~$\rho_0=\rho$ and
\beq \label{firstzero}
\frac{d}{d\tau} \Sact(\rho_\tau) \Big|_{\tau=0} = 0 = \frac{d}{d\tau} \T(\rho_\tau) \Big|_{\tau=0}\:,\qquad
\frac{d}{d\tau} G(\rho_\tau) \Big|_{\tau=0} = 0 \:,
\eeq
the second variation satisfies the inequality
\begin{align*}
\frac{d^2}{d\tau^2} \Sact(\rho_\tau) \Big|_{\tau=0} \geq &- c \; \bigg\| \frac{d^2}{d\tau^2} G(\rho_\tau)
\Big|_{\tau=0}\, \bigg\|_{\R^L} \\[0.2em]
&- \left\{ \begin{array}{ll} 0 & \text{if~$\T(\rho)<C$} \\[0.5em]
\displaystyle c \,\max \left( 0, \frac{d^2}{d\tau^2} \T(\rho_\tau) \Big|_{\tau=0} \right)
& \text{if~$\T(\rho)=C$} \:. \end{array} \right. 
\end{align*}
\end{Prp}
\Proof We consider similar to~\eqref{trho} the variation
\[ \tilde{\rho}_\tau = \Phi \big( \rho_\tau, h(\rho_\tau, \kappa \tau^2), \kappa \tau^2 \big) \:. \]
From~\eqref{Gsat} one sees that the linear constraints are satisfied.
Moreover, a short calculation using~\eqref{firstzero} shows that the first variation of~$\T$ vanishes,
and that by choosing~$\kappa$ sufficiently
large, one can arrange that the second variation of~$\T$ becomes negative.
Now we can argue just as in the proof of Proposition~\ref{prplagrange}.
\QED

\subsection{First Variations with Fixed Support} \label{sec32}
We want to apply Proposition~\ref{prplagrange} to specific variations $(\rho_\tau)_{\tau \in [0,1]}$.
Here we begin with variations keeping the support of~$\m$ fixed, i.e.
\[ \supp \tilde{\m}_\tau = \supp \m \qquad \text{for all~$\tau$}\:. \]
It turns out that it is most convenient to work in the formalism of moment measures introduced in
Section~\ref{mommeas}. In view of~\eqref{RN} and Proposition~\ref{momentprop},
the moment measures corresponding to any minimizer~$\rho \in \M$ are uniquely
characterized by a normalized positive regular Borel measure~$\m^{(0)}$ on~$\K$ and a
function~$f \in L^2(\K, d\m^{(0)})$, being odd in the sense of~\eqref{fodd}.
Conversely, given any positive regular Borel measure~$\m^{(0)}$ and
any function~$f \in L^2(\K, d\m^{(0)})$ (which need not necessarily be odd),
we can define a measure~$\rho \in \M \cap \B$ by~\eqref{rhorep}.
For ease in notation, we will often omit the superscript~$^{(0)}$.
On~$\K$ we introduce the functions
\begin{align}
\ell(x) &= f(x)^2 \int_\K \L(x,y)\: f(y)^2\: d\m(y) && \hspace*{-2cm} \in L^1(\K, d\m) \label{elldef} \\
\mathfrak{t}(x) &= f(x)^2 \int_\K |A_{xy}|\: f(y)^2\: d\m(y)  && \hspace*{-2cm} \in L^1(\K, d\m) \label{tdef} \\
g_l(x) &= f(x) \:\Tr (e_l \, x )\:,\quad l=1,\ldots, L\:,  && \hspace*{-2cm} \in L^2(\K, d\m) \:, \label{gldef}
\end{align}
where~$(e_1, \ldots, e_L)$ again denotes the basis of the symmetric $k \times k$-matrices
used in~\eqref{G1}.
Comparing with~\eqref{twomom0}, \eqref{twomom} and~\eqref{G1},
one sees that integrating over~$x$ with respect to~$d\m$ gives (up to the irrelevant additive
constants~$\Tr(e_l)$ in~$G_l$) the functionals denoted by the corresponding capital letters.
Moreover, we denote the constant function one on~$\K$ by~$1_\K$.
We denote the scalar product on~$L^2(\K, d\rho)$ by~$\la .|. \ra$.

\begin{Lemma} \label{lemmalt}
Under the assumptions of Proposition~\ref{prplagrange}, there are constants~$\kappa, c \in \R$
such that
\beq \label{lt}
\ell(x) + \kappa\, \mathfrak{t}(x) = c \qquad \text{on~$\supp \m$}\:.
\eeq
\end{Lemma}
\Proof Assume conversely that the statement is false. Then there is a set~$\Omega \subset \K$
of positive measure such that on~$\Omega$, the function~$\ell$ is not a linear combination
of~$\mathfrak{t}$ and~$1_\K$, and moreover the restrictions~$\ell|_\Omega$
and~$\mathfrak{t}|_\Omega$ are bounded functions.
Then~$\ell|_\Omega$ is not in the span of the vectors~$\mathfrak{t}|_\Omega, 1_\Omega \in L^2(\Omega, d\m)$.
By projecting~$\ell|_\Omega$ onto the orthogonal complement of these vectors, we
obtain a bounded function~$\psi \in L^\infty(\Omega, d\m)$ such that
\beq \label{psirel}
\la \psi | \ell \ra < 0 \qquad \text{but} \qquad
\la \psi | \mathfrak{t} \ra = 0 = \la \psi | 1_\K \ra \:.
\eeq
Extending~$\psi$ by zero to~$\K$, these relations again hold and~$\psi \in L^\infty(\K, d\m)$.

We now consider the variation of the moment measures
\beq \label{mvar}
d\tilde{\m}_\tau = (1-\tau \psi)\, d\m \:,\qquad \tilde{f}_\tau = (1+\tau \psi)\, f \:,\qquad
\tau \in (-\varepsilon, \varepsilon)\:.
\eeq
The last equation in~\eqref{psirel} implies that~$\tilde{\m}$ is normalized, also
it is positive measure for sufficiently small~$\varepsilon$.
A direct computation using~\eqref{psirel} gives
\[ \frac{d}{d \tau} G_l(\rho_\tau) \big|_{\tau=0} = 0 \:,\quad
\frac{d}{d \tau} \T(\rho_\tau) \big|_{\tau=0} = 2\, \la \psi | \mathfrak{t} \ra = 0 \:,\quad
\frac{d}{d \tau} \Sact(\rho_\tau) \big|_{\tau=0} = 2\, \la \psi | \ell \ra < 0 \:. \]
Hence the first variation decreases the action without changing the constraints.
This is a contradiction to Proposition~\ref{prplagrange}.
\QED

\begin{Lemma} The parameter~$\kappa$ in Lemma~\ref{lemmalt} can be chosen to be
non-negative.
\end{Lemma}
\Proof If the function~$\ell$ is constant, we can choose~$\kappa=0$. Otherwise,
as in the proof of Lemma~\ref{lemmalt} we can choose a function~$\psi \in L^\infty(\K, d\m)$
such that
\[ \la \psi | 1_\K \ra = 0 \qquad \text{and} \qquad \la \psi | \ell \ra = -1\:. \]
Then~\eqref{lt} implies that
\[ \kappa\, \la \psi | \mathfrak{t} \ra = - \la \psi | \ell \ra = 1\:. \]

If~$\kappa$ were negative, by~\eqref{mvar} we could vary the measure~$\rho$ in~$\M \cap \B$
such that the first variation decreases both~$\Sact$ and~$\T$. 
This is a contradiction to Proposition~\ref{prplagrange}.
\QED

\begin{Lemma} \label{lemma39}
Under the assumptions of Proposition~\ref{prplagrange}, there
are real parameters $\lambda_1, \ldots, \lambda_L$ such that
\beq \label{fcond}
\sum_{l=1}^L \lambda_l \, g_l = 4 \,(\Sact + \kappa \T) \:1_\K \qquad \text{on~$\supp \m$}\:.
\eeq
\end{Lemma}
\Proof We first want to prove that~$g_1$ lies in the span of the other functions,
\beq \label{lindep}
g_1 \in \bra 1_\K, g_2, \ldots, g_L \ket \:.
\eeq
If this were not true, just as in the proof of Lemma~\ref{lemmalt}, we could find a
function~$\psi \in  L^\infty(\K, d\m)$ such that
\[  \la \psi | g_1 \ket = k \qquad \text{and} \qquad
\la \psi | 1_\K \ra = 0 = \la \psi | g_l \ra \:,\quad l=2,\ldots, L\:. \]
We consider the variation of the moment measures
\[ d\tilde{\m}_\tau = (1+2\tau \psi)\, d\m \:,\qquad \tilde{f}_\tau = (1-\tau \psi - \tau)\, f \:,\qquad
\tau \in (-\varepsilon, \varepsilon)\:. \]
From our choice of the matrices~$e_l$ (see after~\eqref{G1}), we know that
\beq \label{gint}
\int_\K g_1\: d\m = k \qquad \text{and} \qquad
\int_\K g_l\: d\m = 0 \quad \text{for~$l=2,\ldots, L$}\:.
\eeq
A direct computation yields
\[ \frac{d}{d \tau} G_l(\rho_\tau) \big|_{\tau=0} = 0
\:,\quad
\frac{d}{d \tau} \T(\rho_\tau) \big|_{\tau=0} = -4 \T(\rho)\:,
\quad
\frac{d}{d \tau} \Sact(\rho_\tau) \big|_{\tau=0} = -4 \Sact(\rho)\:. \]
Thus the first variation decreases both~$\Sact$ and~$\T$ without changing the linear constraints.
This is a contradiction, thereby proving~\eqref{lindep}.

According to~\eqref{lindep}, there are real coefficients~$c$ and~$\lambda_2, \ldots, \lambda_L$ such that
\[ g_1 = c \,1_\K + \sum_{l=2}^L \lambda_l g_l \:. \]
Integrating over~$\K$ and using~\eqref{gint}, we find that~$k = c$.
Hence~$c$ is non-zero, and rescaling the~$\lambda_l$ gives the result.
\QED
Combining the results of the previous lemmas, we obtain the following result.
\begin{Thm} \label{thm311}
Assume that~$\rho$ is a minimizer of the 
variational principle of Definition~\ref{defcausal}, where the constant~$C$ satisfies~\eqref{Cbound}.
Then there are Lagrange multipliers~$\kappa \geq 0$
and~$\lambda_1, \ldots, \lambda_L \in \R$ such that for almost all~$x \in \supp \m \subset \K$,
the following identities hold,
\beq \label{doppel}
\frac{1}{4} \sum_{l=1}^L \lambda_l \,g_l(x) = \Sact + \kappa \T = \ell(x) + \kappa \mathfrak{t}(x) \:.
\eeq
In the case~$\T(\rho) < C$, we may choose~$\kappa=0$.
\end{Thm} \noindent
Setting
\beq \label{Lambdadef}
\Lambda = \sum_{l=1}^L \lambda_l \,e_l \:,
\eeq
using~\eqref{gldef} and rewriting the first equation in~\eqref{doppel} in terms of the measure~$\rho$
yields Theorem~\ref{thmhyper}.

\subsection{First Variations with Varying Support} \label{sec321}
We now consider first variations which change the support of the measure~$\m$.
The following notion turns out to be helpful.
\begin{Def} A minimizing measure~$\rho$ is called {\bf{regular}} if the following two conditions are satisfied:
\begin{itemize}
\item[(1)] In the case of the identity constraint~(IC), the functions~$g_1, \ldots, g_L$ must be
linearly independent.
\item[(2)] When~$\T(\rho)=C$, the function~$\mathfrak{t}$ must be non-constant on~$\supp \m$.
\end{itemize}
If one of these conditions is violated, $\rho$ is called {\bf{singular}}.
\end{Def} \noindent
Note that in the case of the trace constraint~(TC), we know from the first equation in~\eqref{gint}
that the function~$g_1$ is non-zero, so that the functions~$g_1, \ldots, g_L$ are automatically
linearly independent. It is an open problem if or under which assumptions all minimizers are regular.

We first analyze regular minimizers (for singular minimizers see Theorem~\ref{thm3}
below). Recall that, according to Theorem~\ref{thmhyper}
the function~$\Phi$ defined by~\eqref{Phidef} (with~$\Lambda$ again given by~\eqref{Lambdadef})
is constant on the support of~$\rho$.
The following result shows that~$\Phi$ is minimal on the support of~$\rho$.
\begin{Thm} \label{thm2} Assume that~$\rho$ is a regular minimizer of the causal
variational principle of Definition~\ref{defcausal}, where the constant~$C$ satisfies~\eqref{Cbound}.
Then
\[ \Phi(x) \geq -2 \left( \Sact + \kappa \T \right) \qquad \text{for all~$x \in \F$} \:. \]
\end{Thm}
\Proof We first consider a point~$x_0 \in \supp \m$. Then we know from Theorem~\ref{thmhyper} that
\[ \Phi(t x_0)\big|_{t=f(x_0)} = -2 (\Sact +\kappa \T) \qquad \text{and} \qquad
\frac{d}{dt} \Phi(t x_0)\big|_{t=f(x_0)} = 0 \:. \]
As~$\Phi(t x_0)$ is a quadratic polynomial in~$t$ with a non-negative quadratic term,
it follows that~$\Phi(t x_0)$ is minimal at~$t=f(x_0)$.

Next we choose~$x_0 \in \K \setminus \supp \m$.
For given~$f_0 \in \R$ and~$\psi, \phi \in L^\infty(\K, d\m)$ with
\beq \label{prod1}
\la \phi | 1_\K \ra = 1 \:,
\eeq
we consider the variation
\begin{align*}
\tilde{\m}_\tau &= (1-\tau \,\phi)\, \m + \tau \delta_{x_0} \\
\tilde{f}_\tau(x) &= \left\{ \begin{array}{ll} \big( 1+\tau \psi(x) +\tau \phi(x) \big) f(x) & \text{if~$x \in \supp \m$} \\[0.5em]
f_0 & \text{if~$x=x_0$}\:. \end{array} \right.
\end{align*}
The first variation is computed by
\begin{align}
\frac{d}{d\tau} G_l \big|_{\tau=0} &= g_l(x_0) + \int_\K \psi\, g_l\, d\m \label{Glvar} \\
\frac{d}{d\tau} \T \big|_{\tau=0} &= 2\, \mathfrak{t}(x_0) + 2 \int_\K (2 \psi + \phi)\, \mathfrak{t}\, d\m
\label{Tvar} \\
\frac{d}{d\tau} \Sact \big|_{\tau=0} &= 2\, \ell(x_0) + 2 \int_\K (2 \psi + \phi)\, \ell\, d\m
\end{align}
(with~$\ell(x_0)$, ${\mathfrak{t}}(x_0)$ and~$g_l(x_0)$ as defined by~\eqref{elldef}--\eqref{gldef}).
Since the functions~$g_l$ are linearly independent, we can choose~$\psi$ such that~$\partial_\tau G_l=0$
for all~$l=1,\ldots, L$. Multiplying~\eqref{Glvar} by~$\lambda_l$ and summing over~$l$,
we can apply Lemma~\ref{lemma39} to obtain
\beq \label{prod2}
4 \,(\Sact + \kappa \T) \:\la \psi | 1 \ra = -\sum_{l=1}^L \lambda_l \:g_l(x_0)\:.
\eeq
Next, using that the function~${\mathfrak{t}}$ is not constant, we
can choose~$\phi$ such that~$\partial_\tau \T=0$. Applying Proposition~\ref{prplagrange}, we
conclude that~$\partial_\tau \Sact \geq 0$. Hence, again using the fact that~$\partial_\tau \T=0$,
we obtain
\begin{align*}
0 &\leq \frac{1}{2}\: \frac{d}{d\tau} \left( \Sact + \kappa \T \right) \big|_{\tau=0}
= (\ell + \kappa {\mathfrak{t}})(x_0) + \int_\K (2 \psi + \phi)\, (\ell + \kappa {\mathfrak{t}})\: d\m \\
&\!\!\!\overset{\eqref{doppel}}{=} (\ell + \kappa {\mathfrak{t}})(x_0) + 
(\Sact + \kappa \T)\: \big\la 2 \psi + \phi \big| 1 \big\ra \:.
\end{align*}
Using~\eqref{prod1} and~\eqref{prod2}, we obtain
\[ (\ell + \kappa {\mathfrak{t}})(x_0) + (\Sact + \kappa \T)
- \frac{1}{2} \sum_{l=1}^L \lambda_l \:g_l(x_0) \geq 0 \:. \]
Applying~\eqref{Lambdadef} and rewriting the resulting inequality on~$\F$ gives the result.
\QED

For singular minimizers the following weaker statement holds.
\begin{Thm} \label{thm3}
Assume that~$\rho$ is a singular minimizer of the 
variational principle of Definition~\ref{defcausal}, where the constant~$C$ satisfies~\eqref{Cbound}.
Let~$\P \subset \F$ be the set
\begin{align*}
\P &= \Big\{ x \in \F \:|\: \text{ there exist } \phi, \psi \in L^1(\K, d\m)
\text{ with } \la \phi | 1 \ra =1 \:, \\
&\qquad g_l(x_0) = -\int_\K \psi \,g_l\: d\m \qquad \text{and} \qquad
\mathfrak{t}(x_0) = -\int_\K (2\psi + \phi) \,\mathfrak{t}\: d\m \Big\}\:,
\end{align*}
where we set~$x_0 = x/\|x\| \in \K$ and~$f(x_0) = \|x\|$.
Then
\[ \Phi(x) \geq -2 \left( \Sact + \kappa \T \right) \qquad \text{for all~$x \in \P$} \:. \]
\end{Thm}
\Proof If~$x \in \P$, we can clearly arrange that~\eqref{Glvar} and~\eqref{Tvar} vanish.
Now we can proceed exactly as in the proof of Theorem~\ref{thm2}.
\QED
We point out that if~$x \in \supp \rho$, then~$x$ lies in~$\P$, as can be seen by
setting~$x_0=x/\|x\|$ and considering the
series~$\phi_n \rightarrow \delta_{x_0}$, $\psi_n \rightarrow -\delta_{x_0}$.
We also remark that if the function~${\mathfrak{t}}$ is {\em{not}} constant, then the condition for~$\mathfrak{t}(x)$
in the definition of~$\P$ can clearly be satisfied. Thus in this case, $\P$ is defined by linear relations,
thereby making it into the intersection of~$\F \subset \Lin(\H)$ with a plane through the origin.

\subsection{Second Variations with Fixed Support} \label{sec34}
For the analysis of second variations, we shall use spectral methods. To this end, 
we use the abbreviations
\begin{align}
\L_\text{eff}(x,y) &= \L(x,y) + \kappa\, |A_{xy}|^2 \label{Leffdef} \\
L(x,y) &= \big( \L(x,y) + \kappa\, |A_{xy}|^2 \big) f(x)^2\: f(y)^2\:. \label{Lxydef}
\end{align}
Then the second equation in~\eqref{doppel} can be expressed as
\beq \label{Lconst}
f(x)^2 \int_\K \L_\text{eff}(x,y)\:f(y)^2\: d\m(y)
\equiv \int_\K L(x,y)\: d\m(y) \equiv \Sact + \kappa \T\:.
\eeq
We also consider~$L(x,y)$ as the integral kernel of a corresponding operator
\beq \label{Ldef}
L \::\: L^2(\K, d\m) \rightarrow L^2(\K, d\m) \:,\quad (L \phi)(x) := \int_\K L(x,y)\, \phi(y)\: d\m(y)\:.
\eeq
\begin{Prp} Under the assumptions of Theorem~\ref{thm311},
the operator~$L$ is self-adjoint and Hilbert-Schmidt.
\end{Prp}
\Proof Obviously, the operator~$L$ is formally self-adjoint. Thus it remains to show that the Hilbert-Schmidt norm
is finite. Using~\eqref{Lconst}, we obtain
\begin{align*}
\|L\|_2^2 &= \iint_{\K \times \K} L(x,y)^2\: d\m(x)\: d\m(y) \\
&\leq \iint_{\K \times \K} \esssup_{y' \in \K} L(x,y')\: \esssup_{x' \in \K} L(x', y)\:  d\m(x)\: d\m(y) \\
&= \Big( \int_\K \esssup_{x' \in \K} L(x', y)\: d\m(y) \Big)^2 = (\Sact + \kappa \T)^2\:,
\end{align*}
concluding the proof.
\QED \noindent
We remark that, similar to~\cite[Lemma~1.9]{continuum}, one could prove that
the sup-norm of~$L$ is an eigenvalue of~$L$ with~$1_\K$ as a corresponding eigenvector.
However, it is not clear in general whether this eigenvalue is non-degenerate.

Since every Hilbert-Schmidt operator is compact, we know that~$L$ has a spectral
decomposition with purely discrete eigenvalues and finite-dimensional eigenspaces.

\begin{Thm} \label{thm4} Assume that~$\rho$ is a minimizer of the 
variational principle of Definition~\ref{defcausal}, where the constant~$C$ satisfies~\eqref{Cbound}.
If~$\T(\rho)=C$, we assume furthermore that the function~$\mathfrak{t}$ is not constant on~$\supp \m$.
Then the operator~$L$ is positive semi-definite on the subspace
\[ J := \bra \mathfrak{t}, g_1, \ldots, g_L \ket^\perp \subset L^2(\K, d\m) \:. \]
\end{Thm}
\Proof We consider the operator~$\pi_J L \pi_J$, where~$\pi_J$ is the orthogonal projection onto~$J$.
Assume on the contrary that this operator is not positive semi-definite.
Since this operator is compact,
there is a negative eigenvalue~$\lambda$ with corresponding eigenvector~$v \in L^2(\K, d\m) \cap J$.
Let us show that there is a bounded function~$u \in L^\infty(\K, d\m) \cap J$
with~$\la u | L u \ra < 0$. To this end, we choose a nested sequence of measurable
sets~$A_i \subset \supp \m$ such that~$\m( \K \setminus \cup_i A_i) =0$ and
the functions~$v, \mathfrak{t}, g_1, \ldots, g_L$ are bounded on each~$A_i$
(this is clearly possible by Chebycheff's inequality).
We let~$v_i \in L^2(A_i, d\m)$ be the projection of~$v|_{A_i}$
onto the subspace~$\bra \mathfrak{t}_{|A_i}, g_{1|A_i}, \ldots, g_{L|A_i} \ket^\perp
\subset L^2(A_i, d\m)$. Then the functions~$v_i$ are clearly bounded.
The dominated convergence theorem
shows that~$\la v_i | L v_i \ra \rightarrow \la v | L v \ra < 0$. Hence~$u=v_i$
for sufficiently large~$i$ has the announced properties.

In view of Lemma~\ref{lemma39}, we know that~$\la u | 1_\K \ra=0$.
Next, we choose a function~$\phi \in L^\infty(\K, d\m)$ satisfying
\beq \label{ortho2}
\la \phi | 1_\K \ra = 0  \:.
\eeq
Then the normalization of~$\m$ is preserved by the following variation,
\begin{align*}
\tilde{\m}_\tau &= (1+\tau u - \tau^2 \phi)\, \m \\
\tilde{f}_\tau(x) &= \left( 1+ \tau^2 \phi \right) f(x) \:.
\end{align*}
A straightforward calculation using the orthogonality relations of~$u$ and~$\phi$ yields
\begin{align}
G_l(\tau) &= G_l(0) + \O(\tau^3)  \\
\T(\tau) &= \T(0) + \tau^2\, \la \phi | \mathfrak{t} \ra + \tau^2 \, \la u | T u \ra + \O(\tau^3) \label{Tvar2} \\
(\Sact+ \kappa \T)(\tau) &= (\Sact+\kappa \T)(0) + \tau^2 \, \la u | L u \ra + \O(\tau^3) \:,
\end{align}
where~$T$ is the operator with the integral kernel~$T(x,y) = |A_{xy}|^2 f(x)^2\: f(y)^2$.
Since the function~$\mathfrak{t}$ is not constant, by suitably choosing~$\phi$ we can arrange that
the quadratic term in~\eqref{Tvar2} vanishes.
Moreover, the term~$\la u | L u \ra = \lambda\, \|u\|^2$ is negative.
Thus we have found a variation which preserves the constraints quadratically, but decreases the action.
This is a contradiction to Proposition~\ref{prplagrange2}.
\QED

\subsection{Second Variations with Varying Support} \label{sec35}
In this section we generalize Theorem~\ref{thm4} to the case when the Hilbert space~$L^2(\K, d\m)$
is extended by a one-dimensional vector space consisting of functions supported on a set which is
disjoint from the support of~$\m$. More specifically, we choose a normalized measure~$\n$
on~$\K$ with
\[ \supp \n \cap \supp \m = \varnothing \:. \]
We arbitrarily extend the function~$f$ to~$\supp \n$.

For the analysis of second variations, we introduce the Hilbert space~$({\mathfrak{H}}, \bra .|. \ket)$ as
\[ {\mathfrak{H}} = L^2(\K, d\m) \oplus \R \:. \]
We extend the operator~$L$, \eqref{Ldef}, to~${\mathfrak{H}}$ by
\begin{align*}
L \,(u, a) &= \left( \phi, b \right) \qquad \text{with} \qquad \\
\phi(x) &= \int_\K L(x,y)\, u(y)\, d\m(y) + a \int_\K L(x,y) \,d\n(y) \\
b &= \iint_{\K \times \K} L(x,y)\, u(y)\, d\m(y)\, d\n(x) + a \iint_{\K \times \K} L(x,y)\, d\n(x)\, d\n(y)\:.
\end{align*}
Then the following theorem holds.
\begin{Thm} \label{thm5} Assume that~$\rho$ is a minimizer of the 
variational principle of Definition~\ref{defcausal}, where the constant~$C$ satisfies~\eqref{Cbound}.
If~$\T(\rho)=C$, we assume furthermore that the function~$\mathfrak{t}$ is not constant on~$\supp \m$.
Then the operator~$L$ is positive semi-definite on the subspace
\[ J := \bra \mathfrak{t}, g_1, \ldots, g_L \ket^\perp \subset {\mathfrak{H}} \:. \]
\end{Thm}
\Proof Assume on the contrary that the operator~$\pi_J L \pi_J$ is not positive semi-definite.
Then the operator has a negative eigenvalue~$\lambda$ with corresponding
eigenvector~$v$. Just as in the proof of Theorem~\ref{thm4}, we can choose a
bounded function~$w=(u, a) \in {\mathfrak{H}} \cap J$ with~$\la w | L w \ra <0$.
Possibly by flipping the sign of the function~$w$, we can arrange that~$a \geq 0$.
Next, we again choose a function~$\phi \in {\mathfrak{H}}$ with~$\supp \phi \subset \supp \m$
satisfying~\eqref{ortho2}. Then the variation
\begin{align*}
\tilde{\m}_\tau &= (1+\tau u - \tau^2 \phi)\, \m+ \tau a \n \\
\tilde{f}_\tau(x) &= \left( 1+ \tau^2 \phi \right) f(x)
\end{align*}
is admissible for sufficiently small positive~$\tau$.
Repeating the arguments in the proof of Theorem~\ref{thm4} gives the result.
\QED

\subsection{An A-Priori Estimate} \label{sec36}
We conclude this section with estimates under the additional assumption that
\beq \label{Leffa}
\inf_{x \in \supp \m} \L_\text{eff}(x,x) > 0 \:.
\eeq
This condition is clearly satisfied in the case~$\kappa>0$. In the case~$\kappa=0$,
the estimates in~\cite[Section~4]{discrete} show that~$\L(x,x)$ is bounded from below,
provided that the trace~$\Tr(x)$ is bounded away from zero.
However, it is conceivable that for a general minimizer,~$\Tr(x)$ might have
zeros on the support of~$\rho$, so that~\eqref{Leffa} could be violated.

\begin{Prp} Under the assumptions of Theorem~\ref{thm311} and assuming~\eqref{Leffa},
the function~$f$ is essentially bounded, $f \in L^\infty(\K, d\m)$. 
Moreover, there is a constant~$c=c(\F)$ 
such that for every~$\varepsilon >0$ the inequality
\beq \label{4meps}
\int_\K |f|^{4-\varepsilon}\: d\m \leq \frac{c}{\displaystyle
\inf_{x \in \supp \m} \L_\text{eff}(x,x)}\: \frac{\Sact + \kappa \T}{1-2^{-\varepsilon}} \, 
\eeq
holds.
\end{Prp}
\Proof In order to prove that~$f \in L^\infty(\K, d\m)$, we proceed indirectly and assume 
conversely that~$f$ is not essentially bounded. Then there is
a point~$x \in \K$ such that for every~$\varepsilon>0$,
\beq \label{supess}
\esssup_{B_\varepsilon(x)} |f| = \infty \:.
\eeq
By decreasing~$\varepsilon$, we can arrange by continuity that
\[ \L_\text{eff}(y,z) \geq \delta := \frac{1}{2}\: \inf_{x \in \K} \L_\text{eff}(x,x)
\qquad \text{for all~$y,z \in B_\varepsilon(x)$}\:. \]
Using~\eqref{elldef}, \eqref{tdef} and~\eqref{Leffdef}, we conclude that for any~$y \in B_\varepsilon(x)
\cap \supp \rho$,
\[ (\ell+\kappa \mathfrak{t})(y) \geq f(y)^2 \:\delta \int_{B_\varepsilon(x)} f^2(z)\, d\m(z) \:. \]
The last integral is non-zero in view of~\eqref{supess}.
Thus by choosing~$y$ appropriately, we can make~$(\ell+\kappa \mathfrak{t})(y)$ arbitrarily large, in contradiction
to Theorem~\ref{thm311}.

In order to prove the inequality~\eqref{4meps}, for any~$L>0$ we introduce the set
\[ \K_L = \{ x \in \K \:|\: |f(x)| > L \}\:. \]
Integrating~\eqref{Lconst} over~$\K_L$ gives
\[ \iint_{\K_L \times \K} \L_\text{eff}(x,y)\: f(x)^2\: d\m(x)\:f(y)^2\: d\m(y) = \m(\K_L)\:
(\Sact + \kappa \T) \:. \]
The covering argument in~\cite[Lemma~2.12]{continuum} shows that there is a constant~$c=\delta(\F)>0$ such that
\[  \Big(\int_{\K_L} f^2 \, d\m \Big)^2 \inf_{x \in \K} \L_\text{eff}(x,x) \leq c
\:\m(\K_L)\: (\Sact + \kappa \T) \:. \]
Setting~$c_1 = c / \inf_{x \in \K} \L_\text{eff}(x,x)$, we obtain
\[ L^4\, \m(\K_L)^2 \leq c_1 \: \m(\K_L)\: (\Sact + \kappa \T) \]
and thus
\[ \m(\K_L) \leq c_1\:(\Sact + \kappa \T)\: \frac{1}{L^4}\:. \]
Now we can estimate the integral by considering the sequence~$L_n=2^n$,
\begin{align*}
\int_\K |f^{4-\varepsilon}|\: d\m &\leq \sum_{n=0}^\infty (2 L_n)^{4-\varepsilon}\: \m(\K_{L_n})
\leq c_1\:(\Sact + \kappa \T) \sum_{n=0}^\infty (2 L_n)^{4-\varepsilon}\: L_n^{-4} \\
&\leq 16\,c_1\:(\Sact + \kappa \T) \sum_{n=0}^\infty 2^{-n \varepsilon}
= 16\,c_1\:(\Sact + \kappa \T)\: \frac{1}{1-2^{-\varepsilon}}\:.
\end{align*}
This gives~\eqref{4meps}.
\QED

\section{The Euler-Lagrange Equations in the Equivariant Case} \label{secequi}
In this section we extend the previous results to the setting of a symmetry group
(possibly non-compact) acting on the measures.
To this end, we first replace~$\C^k$ by a Hilbert space~$(\H, \la .|. \ra_\H)$
of possibly infinite dimension. For a given parameter~$n \in \N$,
we again let~$\F \subset \Lin(\H)$ be the set of all operators of rank at most~$2n$
with at most~$n$ positive and at most~$n$ negative eigenvalues.
Moreover, we let~$G$ be a topological group and~$U$ a continuous unitary representation of~$G$
on~$\H$. Then~$G$ also acts on~$\F$ by
\beq \label{UFact}
U(g) \::\: \F \rightarrow \F\:,\quad x \mapsto U(g) \,x\, U(g)^{-1}\:.
\eeq
A Borel measure~$\rho$ on~$\F$ is called {\em{equivariant}} if~$U(g)_* \rho = \rho$ for all~$g \in G$.
An equivariant Borel measure~$\rho$ induces
a measure on the quotient space~$\F/G$. It is called {\em{normalized}} if~$\rho(\F / G) = 1$.
We consider the class of measures
\[ \M_G = \{ \rho \text{ equivariant normalized regular Borel measure on~$\F$} \}\:. \]
We introduce the functionals~$\Sact$ and~$\T$ by
\begin{align}
\Sact &= \int_{\F / G} \int_\F \L[A_{xy}]\: d\rho(x)\, d\rho(y) \label{SG} \\
\T &= \int_{\F / G} \int_\F |A_{xy}|^2\: d\rho(x)\, d\rho(y) \label{TG}
\end{align}
and define the boundedness constraint as before,
\begin{itemize}
\item[(BC)] The {\em{boundedness constraint}}: \qquad\;\;
$\T \leq C$ \\[-0.5em]
\end{itemize}
In place of the trace and identity constraints, we now consider the following linear constraints.
We let~$h_1, \ldots, h_L \in C^0(\F/G, \R)$ be continuous functions which are homogeneous of degree one, i.e.
\[ h_l(\lambda x) = \lambda\, h_l(x) \qquad \text{for all~$x \in \F/G$}\:. \]
For given constants~$\nu_1, \ldots, \nu_L \in \R$ we introduce the functionals
\[ G_l = \nu_l - \int_{\F/G} h_l(x)\: d\rho(x) \:. \]
\begin{itemize}
\item[(LC)] The {\em{linear constraints}}: \quad\;\;
$\displaystyle  G_l = 0 \quad \text{for all} \quad l=1,\ldots, L$.
\end{itemize}
\begin{Def} \label{defequi}
For any parameter~$C > 0$, our {\bf{equivariant causal variational principle}} is to
minimize~$\Sact$ by varying~$\rho \in \M_G$ under the constraints~{\rm{(BC)}} \text{and}~{\rm{(LC)}}.
\end{Def} \noindent
If~$\H$ is finite-dimensional, the existence of minimizers follows immediately by applying the
compactness results in~\cite[Section~2]{continuum}. Moreover, the trace and identity
constraints can be reformulated in terms of~(LC).
In the infinite-dimensional situation, the trace constraint is obviously
again of the form~(LC). For the identity constraint, however, it is in general not clear
how by modding out the group action, the integral over~$\F$ in~(TC) can be rewritten
as an integral over~$\F/G$. Furthermore, when~$\H$ is infinite-dimensional,
there are no general existence results. It is to be expected that minimizers exist only for particular
choices of the symmetry group~$G$ and its unitary representation~$U$
(for a specific result in this direction see~\cite[Theorem~4.2]{continuum}).
For simplicity, here we do not consider questions related to existence of minimizers.
Instead, we simply assume that an equivariant minimizer~$\rho$  is given.
Moreover, we only treat the case where~$\K/G$ is compact. The case when~$\K/G$ is non-compact
remains an open problem which goes beyond the scope of the present work.

Introducing the moment measures again by~\eqref{m0def}--\eqref{m2def},
we can rewrite the action and the constraints in analogy to~\eqref{onemom}--\eqref{twomom} 
and~\eqref{RN} by
\begin{align}
G_l &= \nu_l - \int_{\K/G} g_l \, d\m \qquad \text{where} \qquad
g_l(x) := f(x)\, h_l(x) \\
\Sact(\rho) &= \int_{\K/G} \int_{\K} \L[A_{xy}]\: f(x)^2\, f(y)^2\: d\m(x)\, d\m(y) \label{Sequi} \\
\T(\rho) &= \int_{\K/G} \int_{\K}  |A_{xy}|^2\: f(x)^2\, f(y)^2\: d\m(x)\, d\m(y) \:, \label{Tequi}
\end{align}
where~$f \in L^2(\K/G, d\m)$. Note that the integration range of the
integrals in~\eqref{Sequi} and~\eqref{Tequi} is the non-compact set~$\K$.
The fact that~$\Sact$ and~$\T$ are bounded ensures that the integrals exist.
However, it is not clear whether the functionals~$\Sact$ and~$\T$ are
Fr{\'e}chet differentiable (cf.\ Lemma~\ref{lemmafrechet}).
In order to ensure Fr{\'e}chet differentiability, we impose the following condition.
\begin{Def} The minimizer~$\rho$ is called $\T$-{\bf{bounded}} if
\[ \sup_{x \in \K/G} \;\int_\K  |A_{xy}|^2\: f(y)^2\: d\m(y) < \infty\:. \]
\end{Def} \noindent
By straightforward adaptions of the methods used in Section~\ref{sec3}
one derives the following result.
\begin{Thm} Suppose that~$\rho$ is a $\T$-bounded minimizer of the equivariant variational principle
of Definition~\ref{defequi}. Assume that~$\K/G$ is compact and that
\[  C > C_{\min} := \inf \big\{ \T(\mu) \:|\: \text{$\mu \in \M_G$ satisfies~(LC) } \big\} \:. \]
Then for a suitable choice of the Lagrange multipliers
\[ \kappa \geq 0 \qquad \text{and} \qquad \lambda_1, \ldots, \lambda_L \in \R \:, \]
the measure~$\rho$ is supported on the intersection of the level sets~\eqref{hyper},
where the function~$\Phi_2$ is given by~\eqref{Phi2def} and 
\[ \Phi_1(x) := -\sum_{l=1}^L \lambda_l \,h_l(x)\:. \]
In the case~$\T(\rho) < C$, we may choose~$\kappa=0$.
\end{Thm} \noindent
Theorems~\ref{thm2}, \ref{thm3}, \ref{thm4} and~\ref{thm5} also
hold in the equivariant setting for $\T$-bounded minimizers
if we only replace the Hilbert space~$L^2(\K, d\m)$
by~$L^2(\K/G, d\m)$ and the integrals over~$\K$ by integrals over~$\K/G$.

\Thanks{{{\em{Acknowledgments:}} We would like to thank Heiko von der Mosel and
the referee for helpful comments on the manuscript.}

\providecommand{\bysame}{\leavevmode\hbox to3em{\hrulefill}\thinspace}
\providecommand{\MR}{\relax\ifhmode\unskip\space\fi MR }
\providecommand{\MRhref}[2]{%
  \href{http://www.ams.org/mathscinet-getitem?mr=#1}{#2}
}
\providecommand{\href}[2]{#2}

\end{document}